\documentstyle[11pt,newpasp,twoside,epsf]{article}
\begin{document}

\title{Submillimeter Dust Continuum Studies of Low and High Mass Star Formation}
\author{Yancy L. Shirley, Kaisa E. Mueller, Chadwick H. Young, Neal J. Evans}
\affil{Department of Astronomy, The University of Texas at Austin,
       Austin, Texas 78712--1083}

\begin{abstract}

Studying the physical environments of low mass 
and high mass cores using dust continuum emission 
provides important observational constraints on theoretical models
of star formation.  The motivation and procedure for modeling dust continuum 
emission is reviewed and the results of recent surveys towards 
low mass and high mass star forming regions are compared.

\end{abstract}

\section{Introduction}

	Optically thin dust emission at submm and mm wavelengths is a 
powerful probe of the density and temperature structure of the
outer envelope of protostars.  Models of the
dust continuum emission constrain theoretical predictions of 
the structure of forming protostellar cores.  
The resolution of current submm and mm bolometer arrays 
effectively image the outer envelope on scales of $10^3$ to $10^5$ AU.  
The basic procedures for understanding the density and temperature 
structure are reviewed as well as the need for radiative transfer 
modeling (\S 2).  The density and temperature structure of the
envelopes of low mass and high mass star forming regions are compared (\S 3)
and important systematic effects are discussed (\S 4).

\section{Dust Continuum Emission}

The specific intensity, at impact parameter $b$, 
of optically thin dust continuum emission from a spherical envelope is given by
\begin{equation}
I_{\nu}(b) = \int_{los} B_{\nu}[T_d(s)] d\tau = \frac{4\mu m_H h \nu ^3}{c^2} \int_b^{r_o} 
\frac{\kappa_{\nu}(r) n(r)}{ \exp\left[ \frac{h\nu}{k T_d(r)} \right] - 1  }
\;\; \frac{r dr}{\sqrt{r^2 - b^2}} \;\;,
\end{equation}
where $s$ is a distance along the line-of-sight,
$\mu m_H$ is the mean molecular mass of gas in grams, $r_o$ is the outer radius,
$\kappa_ {\nu }(r)$ is the dust opacity in cm$^2$/gram of gas, $n(r)$ is the gas
particle density in cm$^{-3}$, and $T_d(r)$ is the dust
temperature distribution (see Shirley et al. 2000).  
Generally, this integral must be solved numerically.  However, several
simplifying assumptions provide an analytical solution:
(1) The dust emits in the Rayleigh-Jeans limit ($T_d >> h\nu / k$);
(2) The temperature and density follow single power law distributions
($T_d(r) = T_f (r / r_f)^{-q}$ and $n(r) = n_f (r / r_f)^{-p}$);
(3) The dust opacity is constant along the line of sight ($\kappa_{\nu}(r) = 
\kappa_{\nu}$);
(4) $r_o \rightarrow \infty$.
With these assumptions, the specific intensity can be expressed as
a power law in the impact parameter ($I_{\nu}(b) \propto b^{-m}$) with the 
exponent $m = (p + q) - 1$.  The density power law index, $p$, 
is found by fitting a power law to the observed intensity distribution and assuming
a temperature power law index, $q$, to find $p = m + 1 - q$.

	The analytical solution is not applicable at submm wavelengths (including 1.3mm) 
for several reasons.  The dust temperature fails the Rayleigh-Jeans criterion in 
the outer envelopes of low mass and high mass protostars since the dust temperature
drops below 20 K at large radii (i.e., only $2h\nu /k$ at 1.3mm).
The temperature distribution departs from a single power law in the inner regions of
the envelope where the radiative transfer becomes optically thick at UV to mid-IR wavelengths.  
In low mass protostars, heating from the interstellar radiation field (ISRF) becomes
important in the outer envelope causing the dust temperature rise towards the
outside edge of the core.  The observed specific intensity profile has been 
convolved with a complicated beam pattern, containing multiple sidelobes,
and is further modified by chopping and the
detailed observing procedure (e.g., scanning).  Therefore, to understand 
the density structure, we must model the radiative transfer 
to self-consistently calculate the temperature
distribution, and then simulate the observing mode (beam convolution, chopping, etc.)
to compare with observations.

\begin{figure}
\plotfiddle{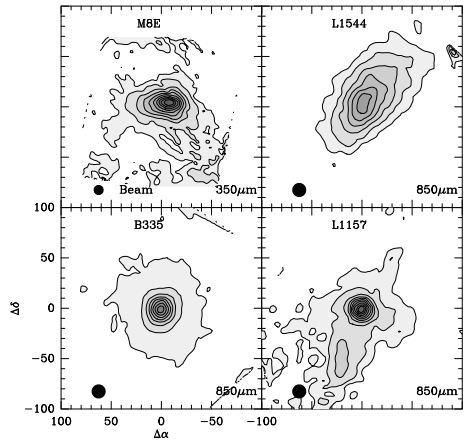}{2.0in}{0}{40}{40}{-120}{-120}
\caption{Submm images of M8E (350 \micron ), L1544 (850 \micron ), B335 
(850 \micron ), and L1157 (850 \micron ) from Mueller et al. (2002) and
Shirley et al. (2000).
}
\end{figure}

\section{Radiative Transfer Models}

Two surveys of the deeply embedded phases of low and high mass
star formation were recently carried out at the University of Texas: 
a SCUBA 850 and 450 $\micron$ survey 
of 39 nearby low mass star forming regions (Shirley et al. 2000, 
Shirley et al. 2002, Young et al. 2002)
and a SHARC 350 micron survey of 51 high mass star forming regions associated with water masers
(Mueller et al. 2002).  Submm images are shown in Figure 1.

The normalized, azimuthally averaged, intensity profiles and spectral
energy distributions (SED) of 19 sources from the SCUBA survey (3 Pre-ProtoStellar
cores, 7 Class 0, and 9 Class I) 
and 31 sources from the SHARC survey were modeled using a
one dimensional radiative transfer code (Egan, Leung, \& Spagna 1988) that takes into
account heating from an internal source, heating from the
ISRF, beam convolution, and
chopping.  The detailed testing of the model parameter space is
discussed in Evans et al. (2001), Shirley et al. (2002), 
Young et al. (2002), and Mueller et al. (2002).  The modeled normalized 
intensity profile is very sensitive 
to the density structure of the core while the modeled SED is sensitive to the
mass and opacity ($\kappa _\nu$).  Ossenkopf \& Henning (1994) opcaities for coagulated
dust grains with thin ice mantles fit the observed SEDs of both samples well.

\begin{figure}
\plotfiddle{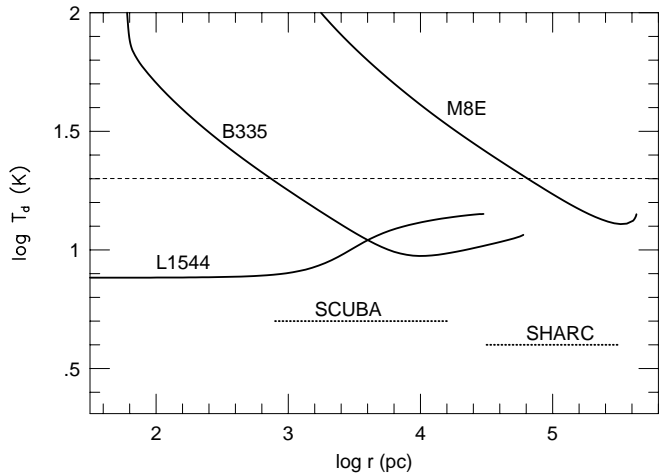}{1.8in}{0}{35}{35}{-140}{-20}
\caption{Temperature profiles from radiative transfer models of L1544 (a low mass PPC), 
B335 (a low mass Class 0 core), and M8E (a high mass core).  The dashed lines show the 
range over which the SCUBA low mass survey and SHARC high mass survey probe the 
envelope for the median distance in each sample ($\theta_{mb}/2$ to $\theta_{chop}/2$).  
A dashed line marks T$_d = 20$ K.
}
\end{figure}

The temperature profile from the best fit radiative transfer models of two 
low mass cores (the PPC L1544 and Class 0 core B335), and a high mass core 
(M8E) are shown in Figure 2.  Single power
law temperature distributions do not fit the calculated profiles in the regions
probed by SCUBA and SHARC.  The temperature profile in a PPC (no internal luminosity
source) drops towards the center as UV to near-IR radiation from the ISRF is attenuated.  
The temperature profile increases dramatically towards the center for
sources with internal luminosity (B335, M8E).  
Heating from the ISRF strongly affects the shape 
of the temperature profile for low mass sources in the region of envelope probed by SCUBA 
(e.g., B335) and has some effect in regions probed by SHARC.

Sources with internal luminosity are well fitted by a single power law density profile.
The histograms of best fit power-law index, $p$, for the low mass and high mass sample
are very similar (Figure 3).  The average $p$ is $1.8 \pm 0.4$ for the high mass
cores and $1.6 \pm 0.4$ for the low mass cores.  The low mass
cores may be sub-divided into Class 0 and Class I objects based on the T$_{bol}$ criterion
(Class 0 typically have T$_{bol} < 70$ K, Chen et al. 1995).  The average $p$ 
for 10 Class 0 cores is $1.7 \pm 0.30$ and
is $1.6 \pm 0.4$ for 9 Class I cores.  No evidence for evolution in the 
shape of the density profile is seen between Class 0 and Class I cores (Young et al. 2002).  

These results can be compared to other submm and mm surveys towards high
mass regions (van der Tak et al. 2000 towards H$_2$O masers
and Hatchell et al. 2000 and Beuther et al. 2002 towards UCHII regions) and
low mass regions (Chandler \& Richer 2000, Hogerheijde et al. 2000, Motte \& Andr\'e 2001,
and J{\o}rgensen et al. 2002).  Beuther et al. (2002) 
observed 69 high mass cores at 1.2 mm and fit broken power laws to the intensity profile.  
The average $p$ is $1.6 \pm 0.5$ in the inner regions ($\theta < 32^{''}$), similar 
to the average $p$ found by Mueller et al. 
There is very little overlap between high mass samples.  The two sources in common
agree withing uncertainties in $p$. 

	The Motte \& Andr\'e (2001)
survey at 1.3 mm towards low mass cores gave an average $p$ steeper by 40\%\ for
10 sources in common.  Motte \& Andr\'e use
the analytical approximation with a single temperature power
law (with q ranging from $-0.2$ to $+0.4$); however, detailed modeling
has shown that the temperature profile changes from falling (positive $q$)
to rising (negative $q$) within the regions of the envelope probed by these
two surveys (Figure 2).  J{\o}rgensen et al. (2002) use a one dimensional model of the
radiative transfer of SCUBA-observed, low mass cores, 
but ignore the effects of the ISRF.  For 7 sources in common, their
average $p$ is flatter by 30\% .  If the ISRF is not included in the model,
the temperature profile will continue to drop towards the outside of the core and 
the resulting best fit density distribution must be flatter to compensate
for the colder dust grains in the outer envelope.  There are significant 
variations between the best fit models from these surveys.  The effects of the ISRF
on the temperature profile can partially explain the differences and must be
included in radiative transfer models (Shirley et al. 2002).

%Uncertainty on p!

\begin{figure}
\plotfiddle{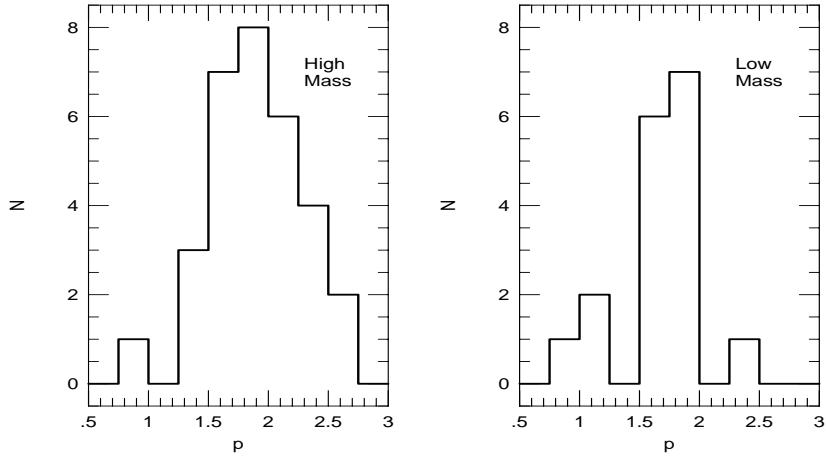}{1.7in}{0}{45}{35}{-180}{-30}
\caption{Histograms of the best fit power law model index for low mass cores
(Shirley et al. 2002 \& Young et al. 2002) and high mass cores (Mueller et al. 2002).
}
\end{figure}

The robustness of the best fit density distribution can be tested by
comparing to the density derived from near-IR extinction
maps. In the case of B335, a low mass
Class 0 protostar, the best fit density distribution 
($n(r) \sim r^{-1.8}$, Shirley et al. 2002) agrees well with the
extinction profile derived from NICMOS images (Harvey et al. 2001)
for radii beyond $5000$ AU.  The near-IR extinction map 
is unable to probe regions of high extinction (A$_{\mathrm{V}}$ $>$ 30 $-$ 50 mag) 
due to a lack of background sources
(Alves et al. 1999, Lada et al. 1999); therefore, we are unable 
to compare methods in the inner regions of the envelope of B335 ($r < 5000$ AU).
Interferometric observations are needed at submm wavelengths to 
test the findings from the dust continuum models at smaller radii.  
Nevertheless, it is encouraging that two different methods are
consistent in the outer region of the envelope.  Further comparisons with
NIR extinction maps are anxiously awaited.

\section{Caveats \& Future Work}

There are many caveats and systematic effects that may affect the
interpretation of the best fit density distribution.  Several sources
(e.g., L1544 and L1157) have asymmetric contours that cannot be
modeled with a one dimensional radiative transfer code.  Multi-dimensional
radiative transfer codes are needed to model asymmetric cores.

Outflows are observed towards many of the sources in our sample.  In 
the near-IR extinction study of B335, it was necessary to consider clearing
of material in outflow cones with a $35^{o} - 45^{o}$ opening angle, while
the submm emission displays no evidence of the outflow (Figure 1).  However,
several cores display emission extending along the outflow direction (e.g., L1157).
A 1D radiative transfer code cannot properly model the effects of the
outflow.

Pure envelope models without disks have been used in the models; however,
disks may contribute a significant fraction of the flux at submm
wavelengths within the central beam.  Since a centrally normalized radial profile is
used, the disk contribution may flatten the interpretation of the
density profile (in an extreme limit) up to $\Delta p \sim -0.5$ 
(Shirley et al. 2002, Young et al. 2002).  Strong constraints on the
disk flux await observations by submm interferometers (SMA and ALMA).  
The potential importance the disk
must not be ignored in future dust continuum studies.

Dust continuum modeling is a powerful diagnostic of the density and temperature
structure of protostellar cores.  The methods of modeling will become more
refined with 3D radiative transport and the inclusion of asymmetries,
outflows, and disks.  
Studies of protostellar envelopes on scales of 10 - 10$^5$ AU will
be possible with the combination of submm interferometers and
single dish bolometer cameras.

%\acknowledgments
%This work has been supported by NASA grants NAG5-7203 and NAG5-10488,
%by NSF grant AST-9988230, and by the State of Texas.

\tiny

\normalsize

\end{document}